\documentclass{PoS}
\usepackage{wrapfig}
\usepackage{algorithm}
\usepackage{algorithmic}
\usepackage{epstopdf}

\title{Recent algorithm and machine developments for lattice QCD}

\ShortTitle{Recent algorithm and machine developments for lattice QCD}

\author{\speaker{Ken-Ichi Ishikawa}\\
        Graduate School of Science, Hiroshima University, Higashi-Hiroshima, Hiroshima 739-8526, Japan.\\
        E-mail: \email{ishikawa@theo.phys.sci.hiroshima-u.ac.jp}}

\abstract{
I review recent machine trends and algorithmic developments for 
dynamical lattice QCD simulations with the HMC algorithm for Wilson-type fermions.
The topics include the trend toward multi-core processors and general purpose GPU (GPGPU) computing, 
and improvements on the quark determinant preconditioning, molecular dynamics integrator, and quark solvers.
I also discuss the prospect on the use of these techniques on the forthcoming petaflops machines.
}

\FullConference{The XXVI International Symposium on Lattice Field Theory\\
		 July 14-19 2008\\
		 Williamsburg, Virginia, USA}

\begin{document}

\section{Introduction}
Lattice QCD is one of the promising method to study the non-perturbative 
aspect of QCD phenomena. 
The real world contains light up and down quarks with a mass in the raneg $2-10$ MeV.  
Since the most difficult part of the lattice QCD algorithm with light quarks 
is computing the quark propagators and incorporating the quark determinant 
to include the quark polarization effects,  
direct simulations at the real quark masses for up and down quarks are difficult.

The algorithm which has been widely used to generate gauge configurations is 
the hybrid Monte Carlo (HMC) algorithm~\cite{Duane:1987de}.
The difficulty in inverting the lattice Dirac opertor needed in the algorithm  
is proportional to the condition number of the operator
and increases by a power of the smallest eigenvalue of the operator.
For Wilson type quarks, since there is no lower bound for the Dirac operator,
the occurrence of small eigenvalues results in a strong violation of energy conservation 
during the molecular dynamics evolution used in the HMC algorithm.
The chiral overlap/domain wall fermions have a lower bound in its spectra, but require
a large computational cost to maintain lattice chirality.
Thus the values of quark masses employed in systematic dynamical QCD simulations 
have been limited to those around the physical strange quark mass ($\sim$50--100 MeV) until 
the end of the last century despite the appearance of the tera-flops supercomputers,
and only staggered-fermions were approaching near the real up-down quark masses.

The situation has changed in the last few years for dynamical lattice QCD simulations.
The technique of preconditioning and separatiopn of low eigenmodes applied to the Dirac operator,  
coupled with the Sexton-Weingarten~\cite{Sexton:1992nu} multiple time step molecular dynamics 
integrator has succcessfully and drastically accelerated the HMC algorithm.
These ideas are rather old~\cite{deForcrand:1996ck}.  
However,  they had to wait recent applications and to reveal their effectiveness at light quark masses. 
Various IR/UV mode separation preconditioners have been developed and investigated.
For optimizing the MD integrator, techniques for tuning the integrator parameters have  
been investigated to optimize the computational cost.

The preconditioning is a basic technique to acceleratre linear equation solvers.
The difficulty to solve linear equations depends on the condition 
number of the coefficient matrix.
The condition number can be reduced by using the spectrum information of the coefficient matrix.
The deflation preconditioning technique, which achieves the reduction by removing 
the low modes, has been investigated and has been known to be a powerful 
preconditioner~\cite{deflWilcox}. 
For the Wilson quark action the low mode deflation based on the local 
coherence assumption significantly improves the solver performance~\cite{deflluscher}.
The multigrid solver/preconditioner can be also a powerful preconditioner
when the effective lattice Dirac operator on coarse grid is adaptively 
constructed. The adaptive multigrid solver, which makes use of 
the same property of the low mode based on the local coherence, has now been applied 
and found to be effective for the Wilson-Dirac quark~\cite{AdaptiveMG}.

Since dynamical lattice QCD simulations demand intensive computing resources,
a variety of supercomputers have been constructed and used.
The recent trend in high performance computing is the use of CPU with 
multiple or many cores on a single chip. 
The bottleneck with multi-core CPU may be the bandwidth between off-chip memory 
and CPU-cores.  The multi-core trend seems to worsens the gap between CPU peak speed and memory bandwidth.
The necessity for parallelism is also enhanced by the multi-core so that 
multi-core CPU's require careful programming for achieving high sustained performance for QCD applications. 
Computing with general Purpose GPU (GPGPU) is an alternate trend which has been recently 
applied to lattice QCD.
The GPU card architecture now available is based on Single Instruction Multiple Data (SIMD) 
or Single Instruction Multiple Thread (SIMT) and can handle thousand of threads and parallelism
with rather high memory bandwidth.

This review is organized as follows. 
We first briefly describe the recent trends of machiens available for lattice QCD simulations,  
centering on multi/many core CPU and GPGPU. 
We then describe the algorithmic developments for dynamical QCD, 
including the preconditioning technique for the QCD partition function and the 
improvements on the molecular dynamics integrator. 
We also discuss mixed precision schemes, the deflation and the multigrid techniques for 
solver improvements. 

Covering all topics related to lattice QCD algorithms is not possible
in a single review. I wish to apologize to those colleagues whose work are not 
reviewed in the paper. 
Here I limit myself to explain only the HMC algorithm with Wilson-type fermions.

\section{Machine trends}

Historically, high performance supercomputers have been developed 
for lattice QCD simulations at each era of the progress of lattice QCD.
The development of these computers involved elaborate tunings 
of various parameters such as memory bandwidth, interconnect bandwidth,
total amount of memory, CPU speed, etc to obtain the best performance and efficiency.
Recent trend shows, however, that the construction of dedicated machines from the CPU chip 
level becomes difficult due to increasing cost of the chip development.
The cluster-type computer built with commodity CPU is an alternative trend 
in high performance computing (HPC).
A guide to build PC clusters for lattice QCD has been described in Ref.~\cite{Holmgren:2004nk}. 
Ref.~\cite{Wettig:2005zc} reviewed not only the PC clusters but also 
commercial supercomputers and the QCD dedicated machines in detail.
New ingredients in the commodity/special parts trend during the recent several years
are the appearance of multi-core CPU, and use of HPC accelerator cards
such as ClearSpeed~\cite{ClearSpeed} or GPGPU.
A detailed benchmark of the QCD kernel performance 
on the recent chip architecture including multi-core and GPGPU 
has been carried out in Ref.~\cite{Ibrahim:2008zz}.

Processor vendors are now producing CPU with two or four cores.
There are several architectures, and most common one is the Intel architecture. 
IBM is supplying CPU's  with PowerPC or Power architecture. 
SUN is also supplying CPU's with the Sparc architecture. 
The trend of commodity CPU chip is multi-core.  
The peak performance is easily multiplied by increasing the number of cores.
The cutting-edge CPU manufacturing process is now around 60-45nm, and 32nm process
will be available in the near future. 
The number of cores will be increased keeping the die size fixed by shrinking 
the process pattern.

The bottleneck of multi-core CPU is the memory bandwidth since 
progress in the speed of memory subsystem is slower than that of the CPU system.
To hide the memory latency between the core and the off-chip memory
an additional memory hierarchy, such as a shared L3 cache memory on the chip,
is inserted between the off-chip memory and the core exclusive cache memory. 
The QCD applications and algorithms should incorporate this architectural pattern 
and trends to keep high sustained speed. 

\paragraph{Commodity PC clusters}
The PC-Cluster type machines based on commodity parts have a better price to peak 
performance ratio. 
The CPU peak speed now achieves 40-50GFlops with the quad core Intel architecture.
However,
commodity architecture is not always suitable for lattice simulations
due to the low memory and interconnect bandwidth,
because the memory bandwidth is still limited to $\sim$ 10 GByte/s and 
a common interconnect is still the Gigabit Ethernet.
The compute intensive part of the lattice QCD simulation is the matrix-vector 
multiplication of the hopping matrix which requires about 3 Byte/Flop in double precision. 
However, a typical value of the Byte/Flop for a common PC with a quad core single CPU running 
at 2.6GHz with a dual channel DDR2-1066 memory is lower than $0.4$.
Moreover the situation becomes worse if the number of core is further increased 
while the memory system is kept the same. 
A similar situation exists in the network bandwidth between compute nodes.
The real efficiency of commodity computing is limited by the bandwidth, and
the total cost for a Cluster machine is determined by the cost of memory bandwidth or
the interconnect bandwidth. 
We can customize the interconnect, or employ Myrinet or Infiniband rather than tre Gigabit 
Ethernet to enhance the network performance, 
but it is difficult to develop a custom-made memory subsystem in view of 
the price to cost performance.

The PC Clusters belong to the category of fat-node supercomputers.

\paragraph{Cell B.E.}
Sony-IBM-Toshiba Cell broadband Engine contains eight-SPU's and one PowerPC PPU
on a single die. This also corresponds to a (heterogeneous) multi-core CPU. 
The first generation Cell B.E. can handle only single precision arithmetic at high speed.
This year, however, IBM released the new generation of Cell B.E. named PowerXCell8i 
with which each SPU has the peak performance 12.8GFlops in full double precision,  
and a total performance of 100GFlops. 
The specific feature of Cell B.E. is that each SPU has its own local store (not cache memory)
and can only access the local store. The data transfer between the local memory 
and off-chip memory is organized by a DMA engine and PPU. 
The bandwidth between the SPU and local memory is kept high, but 
the local to off-chip memory bandwidth is again limited at 25.6 GByte/sec.
The byte per flop is 0.25, so an effective usage of local store 
from algorithmic aspect is required to keep high efficiency.
Using the DMA/PPU and an improved algorithm with a comprehensive control on data flow 
could reduce the bottleneck.

Performance evaluations of lattice QCD with Cell B.E. 
have been reported in Refs.~\cite{Belletti:2007pp,Nakamura:2006zzc,Spray:2008nt},
in which the algorithmic and theoretical aspects have been discussed. 
With the previous Cell B.E., a sustained performace of 40GFlops or 20\% of peak is achieved 
for single precision arithmetic.
A Cell B.E. Cluster named QPACE (QCD PArallel on CEll) is being developed~\cite{QPACE}
aiming at a 200TFlops machine which is based on the new PowerXCell8i CPU's.
This is a fat node type supercomputer and the interconnect bandwidth can be a bottleneck.
To keep a high efficiency for lattice QCD they develop a 3D netwrok with a custom-made network card.

\paragraph{GPGPU}
To achieve high throughput computing with commodity parts, so-called General Purpose 
GPU (GPGPU) computing has been proposed in a couple of years.
Modern graphic processor units (GPU) can compute realistic 3D graphic images
at real time for gaming. 
Although the functions of GPU at early days were limited and fixed for graphics only,
GPU makers are now developing the functionality for more general computing 
including more realistic physical motions
in a virtual world, making use of an increasing availability of transistors per unit area.

The Nvidia and AMD/ATI are the main supplier for the high performance GPU cards.
These companies push GPGPU computing for their GPU products by providing the language and programming 
model, {\i.e., } 
CUDA (Compute unified device architecture)~\cite{CUDAZONE} for Nvidia and 
AMD stream computing for AMD/ATI~\cite{Firestream}.
Recent GPU chips can execute highly parallel operations and threads by
Single Instruction Multiple Data (SIMD) or Single Instruction Multiple Threads (SIMT)
architecture.
They accomplished the peak performance exceeding 1 TFLOPS for single precision on 
a single accelerator. The double precision performance is now at around 100 GFlops.
The bandwidth between GPU and on-board memory is much higher 
than that of the commodity PC's in order to keep the real time graphic computing.
The bandwidth is $\sim$ 100 GByte/sec compared to 16 GByte of dual channel DDR2-1066.
The byte/flop is $\sim$ 1 for double precision and $\sim$ 0.1 for single precision.
The drawback of the GPU computing is less flexibility in the control flow or branch instruction
due to the SIMD like hardware structure. 

If this kind of GPU computing becomes common, a compute node can have a 
peak performance of the order of 1 TFlops. The GPU type cluster also corresponds to a 
fat node supercomputer.
The bottleneck of GPU computing is the bandwidth between host PC and GPU card,
which is usually connected by an PCI-Express bus, and the node-node interconnect bandwidth.
The sustained bandwidth for PCI-E (2nd generation) is about 2GByte/sec. 
To keep high efficiency with the GPU nodes, we have to use an algorithm which sends a large task to 
the GPU card thereby reducing the host-GPU communications.

The first GPGPU computing for lattice QCD has been made in Ref.~\cite{GPGPUQCD} 
using the OpenGL (an API for graphic application)
programming model. A GPU cluster has also constructed, and has been used for 
finite temperature lattice QCD simulations~\cite{Fodor:2007ue}.
The applicability of CUDA for lattice simulations has also been 
reported this year~\cite{RenzoCUDA,RebbiCUDA}.
Ref.~\cite{RebbiCUDA} reports that they achieved over 90 GFlops for the even/odd-site preconditioned
Wilson-Dirac operator in single precision, and about 80GFlops for Wilson CG algorithm
on $16^3\times T$ and $32^3\times T$ lattices using a newest Nvidia's GPU card GTX280.

The Nvidia distributes CUDA SDK and compiler, and the language is very similar to the C language.
The AMD/ATI also distributes Stream SDK which contains CAL (a low level GPU language) and Brook+ 
(a stream computing language and an extension of C++).
These SDK are available from their web pages. 
It seems that Nvidia has an advantage over AMD/ATI because the newbie instruction and 
sample programs for CUDA are widely available on the Internet.
One can easily experiment the CUDA programming with a common PC with Nvidia's 
graphic card (GeForce 8 series and newer).
The cutting-edge HPC card of Nvidia named Tesla series, 
which has no graphic output facility and dedicated for HPC computing,
is also available~\cite{TESLA}.
The AMD/ATI is also shipping HPC dedicated cards named Firestream series~\cite{Firestream}.

\paragraph{$O(1)\sim O(10)$ Peta Flops machine}
PFlops-scale machines can be categorized by the size or the performance of the single compute node.
Thin-node supercomputers, such as BlueGene/L/P and QCDOC, are constructed with 
nodes having a few GFlops of computing performance and a balanced memory and interconnect bandwidth.
This year the Pet-APE (Peta flops Array Processor Experiment) project (successor of the APE series)
has been announced~\cite{PetAPE}. This machine belongs the thin-node supercomputer.
A 1 PFlops machine can be constructed with 100,000 of compute nodes with 10GFlops/node.
This requires a uniform fine grained parallelization to achieve the best performance.
The programing model experienced with lattice QCD is still applicable to the thin-node model.

The fat-node model is based on the node with multiple CPU's and multi-core CPU.
Multi-core CPU's at a 100 GFlops level is already available, 
and higher ones will appear soon. 
Roadrunner~\cite{Roadrunner}, 
which belong to the fat-node model, became the first supercomputer to achieve  
a sustained speed of over 1PFlops~\cite{TopFH} in the world.  
This system is a hybrid supercomputer whose node consists of two dual-core Opteron,  
each attached with an accelerator card containing two PoweXCell8i chips.
The total of 3,456 nodes with 436GFlops of peak speed for each nodes reach 1.5PFlops.
We expect that machines similar to Roadrunner system become more common in near future.
We observe that Roadrunner has a multiple hierarchy for the data flow, {\i.e.,} CPU to CPU, 
off-chip memory to CPU, core-to-core etc., in addition to the node-to-node 
interconnect. A careful programming and a dedicated algorithm
are required on the data flow management for such architecture 
to achieve an efficient performance in this case, such as cache blocking 
within each core to reduce core-to-core data flow in addition to a CPU level data blocking.

\section{Algorithmic developments for dynamical lattice QCD simulations}
\label{sec:Algorithm}

The HMC algorithm has been improved over a long time in order to approach 
physically light quark masses, which may be called the deep chiral region.
The significant improvement in recent years came from the combination of 
preconditioning to split the quark determinant into an IR and UV modes
and using the Sexton-Weingarten~\cite{Sexton:1992nu} multiple-time step molecular dynamics (MTSMD)
integrator to reduce the frequency of Dirac matrix inversions~\cite{deForcrand:1996ck}.

\subsection{Effective action preconditioning and multiple time scale MD integrator}
Preconditioning is a common technique to solve large-scale linear equations efficiently.
Various preconditionings have been used in lattice QCD, such as ILU~\cite{Oyanagi}, 
even/odd site, SSOR~\cite{SSOR} etc..
This technique can be applied to precondition the quark determinant of 
the HMC partition function and has been used to enhance the performance of the HMC algorthm.
        
The lattice QCD partition function contains the quark determinant and 
the computationally dominant part is how to evaluate the quark determinant.
The HMC algorithm introduces a fictitious gauge momentum into the partition function:
\begin{equation}
   {\cal Z}=\int{\cal D}\Pi{\cal D}U \left|\det\left[D\right]\right|^2 e^{-\frac{1}{2}\Pi^2-S_g[U]},
\end{equation}
where $S_g[U]$ is the gauge action, $D$ is the lattice Dirac operator, and $P$ is 
the canonical momenta of the link field $U$.  Here a degenerate $N_f=2$ quarks are assumed.
The configuration $U$ is generated according to the provability weight 
$\left|\det\left[D\right]\right|^2 e^{-\frac{1}{2}\Pi^2-S_g[U]}$. 
The usual HMC algorithm introduces an auxiliary field $\phi$ to evaluate the determinant as
\begin{equation}
   {\cal Z}=\int{\cal D}\Pi{\cal D}U {\cal D}\phi^{\dag} {\cal D}\phi
       e^{-\frac{1}{2}\Pi^2-S_g[U]-|D^{-1}\phi|^2}
     =\int{\cal D}\Pi{\cal D}U {\cal D}\phi^{\dag} {\cal D}\phi
      e^{-H_{\mathit{eff}}[\Pi,U,\phi]},
\end{equation}
where $H_{\mathit{eff}}$ is the effective HMC Hamiltonian.
The HMC algorithm carries out microcanocnical molecular dynamics (MD) evolution with respect to
the effective Hamiltonian followed by a Metropolis accept reject test.
The most time consuming part is the MD evolution, especially the force computation from
the quark potential $|D^{-1}\phi|^{2}$.

The preconditioning technique amounts to a determinant transformation as
\begin{equation}
    \left|\det\left[D\right]\right|^2 = \left|\det\left[PD\right]/\det\left[P\right]\right|^2,
\end{equation}
where $D$ is the lattice Dirac operator, and $P$ is a preconditioner.
Introducing a pseudo fermion fields for each determinant, we have
\begin{equation}
    \left|\det\left[D\right]\right|^2 = 
    \int {\cal D}\phi^{\dag} {\cal D}\phi {\cal D}\chi^{\dag} {\cal D}\chi 
          e^{-\left|(PD)^{-1}\phi \right|^2-\left|P\chi \right|^2}
    = \int {\cal D}\phi^{\dag} {\cal D}\phi {\cal D}\chi^{\dag} {\cal D}\chi 
.          e^{-V_{IR}[U,\phi]-V_{UV}[U,\chi]},
\end{equation}
where $V_{IR}[U,\phi]=\left|(PD)^{-1}\phi \right|^2$ and
$V_{UV}[U,\phi]=\left|P\chi \right|^2$.
If we construct $P$ so that $DP$ contains the IR modes of $D$, and $P$ carries the UV modes of $D$,
the HMC effective action is split into two parts governed by different physical scales.
If, furthermore, the MD force contributions have a hierarchy such that 
$|\delta V_{IR}/\delta U| < |\delta V_{UV}/\delta U|$, a multi time step integration of the 
molecular dynamics (MTSMD) is applicable.
This minimizes the calculation of the cost intensive part $\delta V_{IR}/\delta U$ by assigning
a large time step to the momentum update from $V_{IR}$ and a short time step to the $V_{UV}$ part.
The empirical observation that the large fermionic fluctuation in lattice QCD comes from 
the localized/UV mode allows this splitting and assignment.
This is the recent strategy to improve the HMC algorithm~\cite{deForcrand:1996ck}.

\paragraph{Polynomial preconditioning}
To solve linear equation $Dx=b$, a polynomial of $D$ is a possible 
preconditioner, $P=\sum_{j=0}^{N_{poly}} c_j D^j \sim D^{-1}$.
This preconditioner can be applied to split UV/IR mode to speed up the HMC algorithm 
along with the strategy described above.

Hasenbusch's~\cite{Hasenbusch} heavy mass preconditioner corresponds to 
this category in which $P=(D')^{-1}$ with a heavier quark mass operator $D'$ is chosen.
With this preconditioner
$DP$ contains the light modes and $DP\sim 1$ when the mass difference is small,
while $P^{-1}=D'$ contains the heavyer modes.  Thus the heavy mass parameter works as a UV cut-off.
The original aim for this preconditioner was to extend the MD step size by reducing 
the MD force norm by the proposed determinant separation. Soon after, however, it was realized 
that the combination with the multi time step MD integration leads to an improvement of  
the HMC algorithm~\cite{Urbach:2005ji,AliKhan:2003br,Gockeler:2007rm}.
By tuning the MTSMD step size and the heavy mass of the preconditioner 
$O(10)$ improvement has been observed~\cite{Urbach:2005ji}.
The Hasenbusch's heavy mass preconditioner is applicable to any type of quark action 
and easy to implement. Thus the method has been employed in large scale simulations 
by many groups.

In Ref.~\cite{Peardon:2002wb} a polynomial preconditioner 
$P=\sum_{j=0}^{N_{poly}} c_j D^j \sim D^{-1}$ has been applied to the HMC algorithm.
In this case the order of the polynomial $N_{poly}$ corresponds to the UV cut-off 
instead of the heavy mass parameter of the Hasenbusch's preconditioner.
 
A partial integration of the UV part is possible via the UV-filtering~\cite{Alexandrou:1999ii}.
This filtering has been proposed for the MultiBoson algorithm~\cite{Luscher:1993xx}, but 
it is also applicable to the HMC algorithm~\cite{Ishikawa:2006pb}.
The UV-filter preconditioner is defined by $P\equiv e^{-M}$,
where $M=D-1$. For the unimproved Wilson quark action $M$ is the hopping 
matrix (times $-\kappa$).
Inserting $P$ into the determinant we obtain
\begin{eqnarray}
    \det[D] &=& \det[D e^{-M}]\det[e^{M}], \nonumber\\
            &=& \det[D e^{-M}]e^{{\mathrm Tr}[M]} = \det[D e^{-M}].
\end{eqnarray}
where the trace term is zero for the unimproved Wilson quark.
The remaining matrix $De^{-M}$ can be expanded in $\kappa$ as
\begin{equation}
    De^{-M}= (1+M)e^{-M} = 1 - \frac{M^2}{2} +\frac{M^3}{3}-\frac{M^4}{8}\cdots,
\end{equation}
where the $O(M)$ term is removed and $De^{-M}$ has a smaller condition number. 
Thus the MD force norm is reduced and a larger MD step size is possible~\cite{Ishikawa:2006pb}.
Here we only discussed filtering with the $O(M)$ term.
Higher order filtering and combination with even/odd site preconditioning 
is possible~\cite{Alexandrou:1999ii}.
The UV-filtering is easy to implement only for the quark action with nearest-neighbor coupling;  
otherwise the trace becomes non-local and difficult to implement. 
A further aggressive approach to remove the UV mode from the determinant is 
to change the fermion action to the UV-suppressed one~\cite{UVsuppressF}.

\paragraph{Geometric preconditioning}

For the Wilson and KS type fermions, the most common preconditioner is the even/odd site preconditioning 
using the nearest neigbour nature of the coupling matrix.
By ordering the lattice sites so that even-sites come first and odd-sites the last, 
the Wilson-Dirac operator is expressed as the following block $2\times 2$ matrix:
\begin{equation}
 D = \left(
     \begin{array}{cc}
         1_{ee} & M_{eo} \\
         M_{oe} & 1_{oo} \\
     \end{array}
\right),
\end{equation}
where $1_{ee}$ ($1_{oo}$) is the identity matrix acting on even (odd) sites, and
$M_{eo}$ ($M_{oe}$) is the matrix hopping from odd sites to even sites (vice versa).
Sandwiching with a left $L$ and a right $R$ preconditioner defined by
\begin{equation}
 L =\left(
     \begin{array}{cc}
         1_{ee} & -M_{eo} \\
              0 & 1_{oo} \\
     \end{array}
\right), \ \ \ \ 
 R =\left(
     \begin{array}{cc}
         1_{ee} & 0      \\
        -M_{oe} & 1_{oo} \\
     \end{array}
\right),
\end{equation}
we can diagonalize $D$ in the block form as
\begin{equation}
 LDR = 
\left(
     \begin{array}{cc}
         1_{ee} & -M_{eo} \\
              0 & 1_{oo} \\
     \end{array}
\right)
\left(
     \begin{array}{cc}
         1_{ee} & M_{eo} \\
         M_{oe} & 1_{oo} \\
     \end{array}
\right)
\left(
     \begin{array}{cc}
         1_{ee} & 0      \\
        -M_{oe} & 1_{oo} \\
     \end{array}
\right)=
\left(
     \begin{array}{cc}
         \hat{D}_{ee} & 0      \\
                     0 & 1_{oo} \\
     \end{array}
\right),
\end{equation}
where $\hat{D}_{ee}=1 - M_{eo} M_{oe}$ the Schur complement of $D$. 
Using $\det[R^{-1}]=\det[L^{-1}]=1$ the quark determinant is reduced to
\begin{equation}
    \left|\det\left[D\right]\right|^2 = \left|\det\left[\hat{D}_{ee}\right]\right|^2,
\end{equation}
The pseudo fermion is introduced to the even-site preconditioned
operator $\hat{D}_{ee}$. 
The condition number is reduced and a larger MD step size is possible.
The even/odd-site preconditioned operator $\hat{D}_{ee}$ has been treated as the 
basic operator, and further preconditioning, such as Hasenbusch's one, 
has been applied to $\hat{D}_{ee}$ to enhance the HMC performance.

The ILU preconditioning with global lexicographic ordering is a more powerful 
preconditioner~\cite{Oyanagi} than the even/odd preconditioner.
ILU with various ordering has been investigated~\cite{SSOR}
in the context of linear solver preconditioning.

Application of the ILU preconditioning to the HMC algorithm is also 
possible~\cite{deForcrand:1996ck,Peardon:2000si}.
By ordering lattice sites properly, we can decompose $D$ to $1+U+L$ with
a strictly upper triangular part $U$ and a lower part $L$. 
Using the left and right preconditioning, the quark determinant reduces to
\begin{eqnarray}
    \det[D]&=&\det[(1+L)^{-1}D(1+U)^{-1}]\det[1+L]\det[1+U]\nonumber\\
    &=&\det[(1+L)^{-1}D(1+U)^{-1}]\ \ =\ \ \det[\hat{D}].
\end{eqnarray}
The HMC algorithm based on the determinant $\det[\hat{D}]$ is obtained 
and larger MD step size is expected
when the ILU preconditioned $\hat{D}$ has a reduced condition number.
For linear solver the efficiency of the ILU preconditioning depends on the site ordering
and it is known that there is a trade-off between the efficiency and parallelism~\cite{Ordering}. 
Therefore there may exist a similar the trade-off between the efficiency and parallelism for 
the HMC algorithm with ILU preconditioning.

The geometric preconditioner explained above is based on the site orientation.
A domain oriented preconditioner is also possible as has been 
proposed by L\"{u}scher~\cite{Luscher:2005rx}.
The lattice sites are grouped and colored into two domains, and
the domains are blocked and ordered in a checker board ordering (even/odd ordering).
Based on this domain decomposition, the domain-decomposition HMC (DDHMC) algorithm is constructed.
The preconditioning follows the same way as in the even/odd site preconditioner.
The Wilson-Dirac operator can be represented in a similar block $2\times 2$ form as
\begin{equation}
 D = \left(
     \begin{array}{cc}
         D_{EE} & D_{EO} \\
         D_{OE} & D_{OO} \\
     \end{array}
\right),
\end{equation}
where $D_{EE}$ $(D_{OO})$ is the operator within even-domain (odd-domain),
and $D_{EO}$ $(D_{OE})$ is the operator connecting from odd to even-domain (vice versa).
The determinant reduces to 
\begin{eqnarray}
 \det[D]& =& 
\det\left[
     \begin{array}{cc}
         D_{EE} & D_{EO} \\
         D_{OE} & D_{OO} \\
     \end{array}
\right] \nonumber\\
&=&
\det\left[D_{EE}\right]\det\left[D_{OO}\right]
\det\left[1_{EE}- (D_{EE})^{-1}D_{EO}(D_{OO})^{-1}D_{OE}\right].
\label{eq:DDDET}
\end{eqnarray}
The operator $(1_{EE}- (D_{EE})^{-1}D_{EO}(D_{OO})^{-1}D_{OE})$ is the Schur complement 
by the block even/odd preconditioning.
This preconditioner introduces a sharp cut-off by the block size at which 
the IR and UV scales are decoupled. 
The original proposal is that the block extent $L_{block}$ satisfy 
$\Lambda_{QCD}^{-1} \sim L_{block}$ from the physical point of view.
The Schur complement part represents the physics below the confinement scale and
the domain-restricted part from the lattice cut-off scale to the confinement scale.
Introducing pseudo-fermion fields for each determinant of Eq.~(\ref{eq:DDDET}) we can construct 
the so called domain-decomposed HMC (DDHMC) algorithm.

For actual implementation of the DDHMC further improvements are achieved by preconditioning Eq. (\ref{eq:DDDET})
and by incorporating the dead/alive link method~\cite{Luscher:2005rx}.
The UV part can be preconditioned by the site-oriented even/odd or the ILU preconditioner,  
and the IR part using the spin structure.
The link fields connected to each domain and located parallel to the domain 
surface can be kept fixed during the MD evolution.  With this procedure the internode 
communication for exchanging the updated link fields can be omitted.
This is a semi-local updates algorithm.
To ensure the ergodicity of MCMC, a random shift of the lattice coordinate origin is 
carried out after each HMC evolution.
This method, the dead/alive link method, makes the DDHMC algorithm more effective in 
parallelism and internode bandwidth, but it is applicable only for 
the quark action with nearest-neighbor couplings such as the unimproved/clover Wilson fermion or
simple KS-fermion actions.
The effectiveness of the domain-decomposition factorization strongly depends on 
the structure of the quark action and is maximised for nearest-neighbor coupling action.
 
Extensive studies of the DDHMC algorithm have been carried out in 
Refs.~\cite{DDHMClight,PACSCSDDHMC} which clarified the effectiveness toward the chiral limit.
The efficiency of the DDHMC algorithm generally depends on the shape and the size of the domain.
The 1-dimensional striping domain decomposition version has been investigated in 
Ref.~\cite{Hasenbusch:2007er}.

Another Schur complement approach has been discussed in Ref.~\cite{Borici:2007bp}, where
the RG consideration and the Schur complement approach are combined to construct a better 
kernel for the domain-wall/overlap fermion actions. The derivation shares the same property that
the quark determinant can be transformed using UV/IR separation.

\paragraph{Rational approximation and implicit scale splitting}
Originally the Rational HMC algorithm has been proposed to simulate 
non-local complicated quark actions (this does not mean physically non-local) such as
the two-flavor KS fermion action~\cite{RHMC,RHMC2}, for which a fractional power or 
$n$th-root of the quark determinant is required.
The fractional power decomposition has another benefit to accelerate the HMC algorithm
in that the fractional power reduces the condition number of the quark operator with a 
consequent reduction in the maximum MD force norm as in the case of the 
Hasenbusch's heavy mass preconditioner.
Moreover the rational approximation for the fractional power quark operator introduces 
the possibility to enhance the efficiency through a combination with 
the partial fraction form of the rational 
approximation and the MTSMD integrator. 

Here we explain the method using the two-flavor Wilson quark action as an example.
The quark determinant can be split in $n$ pieces as
\begin{equation}
    |\det[D]|^2 = \det[D^{\dag}D] = \det[Q^2] = \det[(Q^{2})^{\frac{1}{n}}]^{n},
\end{equation}
where $Q\equiv \gamma_5 D$ and $Q^2$ is Hermitian and non-negative. 
By introducing $n$-pseudo fermion fields, we have 
\begin{equation}
    |\det[D]|^2 = \int\left[\prod_{j=1}^{n}{\cal D}\phi_j^{\dag}{\cal D}\phi_j\right]
                \exp\left(-\sum_{j=1}^{n}\phi_j^{\dag}(Q^{2})^{-\frac{1}{n}}\phi_j\right).
\end{equation}
\begin{wrapfigure}{r}{7cm}
\vspace*{-1em}
\includegraphics[width=7cm]{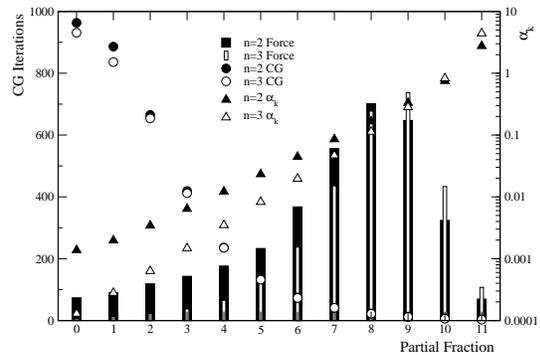}
\vspace*{-2em}
\caption{The variation of the force magnitude (\(L_2\)
    norm), conjugate gradient (CG) cost and \(\alpha_k\) parameter for each
    partial fraction with \(n=2,3\) pseudo-fermions. Figure taken from Ref.~\cite{RHMC2}.}
\label{fig:RHMCForce}
\vspace*{-1em}
\end{wrapfigure}
At this point the (maximum) MD force norm is estimated as $n \mu^{1/n}$ 
($\mu$: the condition number of $Q^2$) naively.
Since the MD force norm without the $n$th-root trick is simply $\mu$, an $n\mu^{1/n-1}$ reduction of 
the MD force norm is expected. 
The optimal $n$ has been estimated as $n_{\mathrm{opt}}=(\log\mu)/2$
via the force norm and the computational cost estimate.

The $n$th-root operator $(Q^2)^{1/n}$ can be well approximated by the rational approximation:
\begin{equation}
 (Q^2)^{1/n}\simeq \sum_{j=1}^{N_{R}}\frac{\alpha_j}{Q^2+\beta_j}.
\end{equation}
This partial fraction form enables us to split the MD potential into several pieces.
The coefficients $\alpha_j$ and $\beta_j$ have the special property, which makes 
the UV/IR splitting effective, such that 
the most cost intensive part coming from the small shift $\beta_j$ has a small 
corresponding residue.  This situation is illustrated in 
Figure~\ref{fig:RHMCForce} which shows the hierarchy of the MD force magnitude as a 
function of fraction number~\cite{RHMC2}.
Making use of this UV/IR property the quark potential $S_q$ can be classified into 
several pieces as
\begin{equation}
    S_q =  \sum_{j=1}^{n}\sum_{k=1}^{N_{cut}}       \alpha_k \phi_j (Q^2+\beta_k)^{-1}\phi_j
         + \sum_{j=1}^{n}\sum_{k=N_{cut}+1}^{N_{R}} \alpha_k \phi_j (Q^2+\beta_k)^{-1}\phi_j,
\end{equation}
where the residue and shift coefficients are ordered to satisfy $\beta_{k} < \beta_{k+1}$.
Here we separate modes into two groups for illustration; the first group corresponds to 
the IR part and the last one the UV part in this case.
The parameter $N_{cut}$ controls the UV/IR separation and is chosen to minimize the total cost.
This mode splitting is rather implicit compared to the determinant preconditioner method.
Combining with the MTSMD integrator the HMC algorithm can be constructed.  The resulting algorithm
is called Rational HMC (RHMC) algorithm.
To make the RHMC algorithm more effective the geometric preconditioning is applied 
to precede the rational approximation in realistic simulations.
The approximation theory for matrix functions including the rational approximation has been 
described in Ref.~\cite{Kennedy:2004tj}.
For more detailed implementation of the RHMC algorithm see Ref.~\cite{Kennedy:2006ax}.

We have discussed the RHMC algorithm for the Wilson quark kernel $Q^2$ for illustration, 
and we did not mention the perfectness of the rational approximation.
To make this algorithm exact we need to know the spectrum boundary of the kernel operator 
($Q^2$ in this example). 
Given the spectrum interval the error from the approximation is easily controlled 
to satisfy a required precision to make the algorithm exact. 
This requires the computation of the smallest and largest eigenvalues for $Q^2$.
In practice, however, to minimize the total computational cost, an accurate rational 
approximation is used only for the Hamiltonian computation and the pseudo-fermion 
field generation, and less accurate approximation is used for the MD evolution~\cite{Kennedy:2006ax}.
If no spectrum interval data are available for the preconditioned matrix, 
we need to put a speculative bound and check the exactness or correct the error with 
reweighting or noisy Metropolis methods with controlled manner~\cite{Kogut:2006jg}.
The RHMC algorithm is applicable to any type of quark action and has been used extensively 
for large scale simulations with various quark 
actions~\cite{Durr:2008rw,Aoki:2005vt,Antonio:2008zz,Gockeler:2007rm,Takaishi:2007ga}.

The rational approximation we have discussed is based on the Hermitian kernel operator $Q^2$.
Non-Hermitian variants are also possible as in the case of the MultiBoson 
and the PHMC 
algorithms~\cite{Borici:1995am,deForcrand:1996ck,Takaishi:2001um,Aoki:2001pt,Peardon:2002wb,Oliver}.
The non-Hermitian variant would especially more effective 
than the Hermitian version in case of the Wilson quark action.
The non-Hermitian rational approximation for $D^{-1/2}$ is obtained via 
the Cauchy's contour integral representation as the matrix function of $D$.
The property of the contour integral representation for a matrix function 
has been investigated in Ref.~\cite{Contourinteg}.

\subsection{Molecular dynamics integrator}

On a par with the determinant preconditioning, the use of Sexton-Weingarten multiple-time 
step MD (MTSMD) integrator is the key technique to enhance the performance of the HMC algorithms.
Here we explain the basics of the MTSMD integrator and the recent improvement on the MTSMD integrator
including the Omelyan-Mryglod-Forlk scheme~\cite{Omelyan,Takaishi:2005tz}
and the tuning method for the MD integrator~\cite{Kennedy:2007cb,HMCPoisson}.
                                                     
The MD integrator used in the HMC algorithms is based on the symplectic 
integrator with exact time-reversal symmetry. 
The classical trajectory $\gamma(t)=(q(t),p(t))$ in the phase space of a system is governed by the 
Hamiltonian with $H(q,p)=T(p)+V(q)$ and the formal solution of the trajectory is written by
\begin{equation}
    \gamma(t) = \exp\left[t \hat{L}_{H} \right] \gamma(0),
\end{equation}
where the linear operator $\hat{L}_{H}$ is defined via the  Poisson bracket $\{,\}$
as $\hat{L}_{H}X \equiv \{ X, H \}$ with $X=X(\gamma)$ a function of $\gamma$.
Although the linear operator $\hat{L}_{H}$ can be decomposed as 
$\hat{L}_{H}=\hat{L}_{T}+\hat{L}_{V}$, the exponential operator cannot be expressed 
in a simple form because $[\hat{L}_{T},\hat{L}_{V}]\ne 0$ in general.
Thus an approximation for the exact evolution operator should be introduced. This
affects the efficiency of the HMC algorithm through violation of the energy conservation.
The MD integrator is usually constructed by decomposing the exponential operator 
$\exp\left[t\hat{L}_{H}\right]$ with the simple operators $Q(t)=\exp\left[ t\hat{L}_{T}\right]$ 
and $P(t)=\exp\left[t\hat{L}_{V}\right]$. 
The operation of $Q$ and $P$ is
\begin{equation}
    Q(t)X(q,p) = X(q + t p,p),\ \ \ \ 
    P(t)X(q,p) = X\left(q,p - t \frac{\partial V}{\partial q}\right).
\end{equation}
For example the second order leapfrog (2LF) integrator is based on the following decomposition:
\begin{equation}
  e^{t\hat{L}_{H}} \simeq
\left(P\left(\frac{\delta t}{2}\right) Q(\delta t) P\left(\frac{\delta t}{t}\right)\right)^{N_{MD}}
\equiv U_{2LF,PQP}(t;N_{MD}),
\end{equation}
with $\delta t = t /N_{MD}$ and $N_{MD}$ the MD step number. 
An alternate version in which the order of $Q$ and $P$ is interchanged is also possible.
Our task is to find a better integrator with higher accuracy and lower computational cost.
The error of the integrator can be evaluated by estimating the so called {\it shadow Hamiltonian} $H'$,
which is close to the original $H$ and is exactly conserved along with the approximate trajectory.
$H'$ can be extracted using the Baker-Campbell-Hausdorff (BCH) formula.
For the 2LF case we define $H'$ via
\begin{equation}
  U_{2LF,PQP}(t,N_{MD}) \equiv e^{\hat{L}_{H'}},
\end{equation}
where we find that $H'=H - \left(\{V,\{V,T\}\}+2\{T,\{V,T\}\}\right) \delta t^2/24 + O(\delta t^4)$.
The 2LF integrator has $\delta t^2$ error.

Higher order integrators can be constructed by nesting the 2LF integrator and tuning 
the step size at each nesting depth to eliminate the $O(\delta t^{2k})$ terms 
in $H'$~\cite{CreutzGockesch,CampostriniRossi,Yoshida,Suzuki}.
It has been shown, however, that the 2LF integrator is the best in efficiency for  
dynamical QCD simulations with light quarks in large volumes~\cite{Takaishi:1999bi}.

Omelyan {\it et al.}\cite{Omelyan} have introduced a new decomposition scheme which minimizes 
the error terms in the shadow Hamiltonian instead of eliminating them.  
This scheme decomposes a single MD step into several steps
and introduces tunable parameters to scale the divided time step.
For example the 2nd order scheme can be further divided to
\begin{equation}
  e^{t\hat{L}_{H}} \simeq
\left(P\left(\lambda \delta t\right) 
      Q\left(\frac{\delta t}{2}\right) 
      P\left((1-2\lambda)\delta t\right)
      Q\left(\frac{\delta t}{2}\right) 
      P\left(\lambda \delta t\right)\right)^{N_{MD}},
\end{equation}
where $\lambda$ is a tunable parameter. The shadow Hamiltonian reads 
\begin{eqnarray}
    H'&=&H+\left[\alpha(\lambda)\{V,\{V,T\}\}+\beta(\lambda)\{T,\{V,T\}\}\right]\delta t^2 + O(\delta t^4), \nonumber\\
   \alpha(\lambda) &=& \frac{6\lambda^2-6\lambda+1}{12},\ \  \beta(\lambda) =  \frac{1-6\lambda}{24},\nonumber
\end{eqnarray}
where $O(\delta t^2)$ terms still remain.
Assuming $\{V,\{V,T\}\}\sim\{T,\{V,T\}\}\sim O(1)$ 
and minimizing $\sqrt{\alpha(\lambda)^2+\beta(\lambda)^2}$,
the $O(\delta t^2)$ error term can be minimized at 
$\lambda_c \simeq 0.193\ 183\ 327\ \cdots$~\cite{Omelyan}. 
This is the so called 2nd order minimum norm integrator (2MN)~\cite{Takaishi:2005tz}. 

\begin{wrapfigure}{r}{7cm}
\vspace*{1em}
\includegraphics[width=7cm]{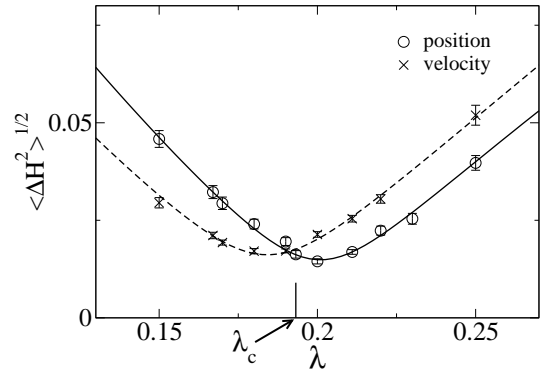}
\caption{$\langle\delta H^2\rangle^{1/2}$  as a function of $\lambda$. 
Simulations are performed at $\beta=5.00$ and $\kappa=0.160$ on  $4^4$ lattices 
with $\delta t=0.05$. Figure taken from Ref.~\cite{Takaishi:2005tz}.}
\label{fig:OptimalTwoMN}
\end{wrapfigure}
This scheme has twice the momentum update compared to the 2LF scheme and
the computational cost is doubled.
However the 2MN integrator would have higher acceptance ratio with smaller violation of 
the energy conservation low so that there is chance to improve the HMC algorithm.
Takaishi and Forcrand~\cite{Takaishi:2005tz} 
have investigated the 2MN integrator and 4th-order minimum norm (4MN) integrator
for the HMC algorithm in lattice QCD.
They found that the 2MN integrator has a better performance than the 2LF integrator 
by about 50\% and the version integrating the position first ($Q$ first) has slightly 
better performance than the reversed version.
The optimal $\lambda$ can be obtained by measuring
$\langle \delta H^2\rangle$ as a function of $\lambda$ since 
$\langle \delta H^2\rangle$ is the metric for the HMC acceptance ratio.
They found that the optimal one may differ from $\lambda_c$ and depends on 
the action and the parameters (Fig.~\ref{fig:OptimalTwoMN}).

The method described above opens a new way to customize the integrator for a given action.
A direct optimization on $\langle \delta H^2\rangle$, however, requires parallel simulations
with several $\lambda$ or MD schemes, which is rather compute intensive.
Clark, Kennedy and Silva~\cite{HMCPoisson} proposed a tuning scheme 
for the MD integrator with less computational cost.
As shown above any MD integrator conserves the corresponding shadow 
Hamiltonian whose error terms are expressed in terms of the Poisson bracket analytically.
Measuring the expectation value of the Poisson brackets on equilibrated configurations 
enables us to reconstruct the error terms as a function of tunable parameters 
for any MD integrator (up to desired order in $\delta t$).
The tuning strategy they proposed is as follows.
Defining the error terms as $\Delta H=H'-H$, the energy difference
after the time evoution with $t$ becomes 
$\delta H\equiv H(t) - H(0) = \Delta H(0) - \Delta H(t)$. 
Thus we expect the optimal acceptance is achieved at the minimum of 
$\langle (\delta H)^2 \rangle = \langle (\Delta H(0) - \Delta H(t))^2 \rangle$.
Assuming that $\Delta H(t)$ and $\Delta H(0)$ are independent and have the same histogram,
we obtain $\langle (\delta H)^2 \rangle = 2 {\mathrm{Variance}}(\Delta H)$ where $\Delta H$ 
are constructed with the Poisson brackets measured on equilibrated configurations.
Therefore we expect that the optimal parameter minimizing the variance of $\Delta H$ 
coincides with that minimizing $\langle (\delta H)^2 \rangle$.
This is a indirect optimization of $\langle (\delta H)^2 \rangle$.
They have demonstrated this method with the 2MN integrators and 
found a good agreement for the optimal $\lambda$ between the direct and the indirect methods.

The Omelyan (or 2MN) integrator can be used as the building block to construct
a recursive MTSMD integrator. In this case the several tunable parameters ($\lambda$ for ex. 2MN) are 
introduced at each depth of the recursion. The motion governed by the Hamiltonian 
with $N_p$ potentials
$H(q,p)=T(p)+\sum_{j=0}^{N_p-1}V_j(q)$, where the potential terms are ordered as 
$|\partial V_{j-1}/\partial q| > |\partial V_j/\partial q|$, can be integrated via
\begin{eqnarray}
    U_{j}(t;\vec{N}_j)&\equiv& \left[
P_{j}\left(\frac{\lambda_j t}{N_j}\right)     U_{j-1}\left(\frac{t}{2 N_j};\vec{N}_{j-1}\right)
P_{j}\left(\frac{(1-2\lambda_j)t}{N_j}\right) U_{j-1}\left(\frac{t}{2 N_j};\vec{N}_{j-1}\right)
P_{j}\left(\frac{\lambda_j t}{N_j}\right)\right]^{N_j},\nonumber\\
    U_{0}(t;\vec{N}_0)&\equiv& \left[
P_{0}\left(\frac{\lambda_0 t}{N_0}\right)     Q\left(\frac{t}{2 N_0}\right)
P_{0}\left(\frac{(1-2\lambda_0)t}{N_0}\right) Q\left(\frac{t}{2 N_0}\right)
P_{0}\left(\frac{\lambda_0 t}{N_0}\right)\right]^{N_0},
\end{eqnarray}
with $\vec{N}_{j}\equiv(N_0,N_1,\cdots,N_j)$ the timescale parameter and 
$P_j(t) X(q,p) \equiv X(q,p-t \partial V_j/\partial q)$ the momentum update with $j$-th potential.
This is based on the PQPQP ordered Omelyan integrator.
The QPQPQ ordered version leads to
\begin{eqnarray}
    &&\hspace*{-1em}U_{j}(t;\vec{N}_j)\equiv\nonumber\\ &&\left[
U_{j-1}\left(\frac{\lambda_j t}{N_j};\vec{N}_{j-1}\right)    P_{j}\left(\frac{t}{2N_j}\right)     
U_{j-1}\left(\frac{(1-2\lambda_j)t}{N_j};\vec{N}_{j-1}\right)P_{j}\left(\frac{t}{2N_j}\right)     
U_{j-1}\left(\frac{\lambda_j t}{N_j};\vec{N}_{j-1}\right) \right]^{N_j},\nonumber\\
   &&
\hspace*{-1em}
U_{0}(t;\vec{N}_0)\equiv \left[
Q\left(\frac{\lambda_0 t}{N_0}\right)    P_{0}\left(\frac{t}{2N_0}\right)
Q\left(\frac{(1-2\lambda_0)t}{N_0}\right)P_{0}\left(\frac{t}{2N_0}\right)
Q\left(\frac{\lambda_0 t}{N_0}\right) \right]^{N_0}.
\end{eqnarray}
Note that the step number to integrate $U_{j-1}$ part can be different 
between the inner and the outer $U_{j-1}$ in the QPQPQ version
since the integration length differs between them and this may further reduce 
the total computational cost. 
The MTSMD 2MN integrator recursively constructed has been used in 
Refs.~\cite{Antonio:2008zz,Boyle:2007fn,Boucaud:2008xu,Durr:2008rw}.

\subsection{Solver improvements}
The performance issue of linear solvers is important in lattice QCD simulations. 
In this section we describe the recent solver improvements for the Wilson-type quark actions.
The recent topics to speed up the solver are categorized in two;
(i)  use of flexible preconditioners with single precision, 
(ii) use of the spectrum information to construct preconditioners.

\paragraph{Flexible preconditioner}
The key for the preconditioner $P$ for solving a linear equation 
is to construct a computationaly inexpensive approximation for $D^{-1} \sim P$. Once $P$ is constructed
the linear equation $Dx = b$ is preconditioned as $DPy=b, x= Py$ or $ PD x = Pb$.
A naive idea for $P$ is the use of  polynomial approximation for $D^{-1}$. However
this does not improve the solver in terms of the total count of matrix-vector multiplication.

If $P$ is computed in single precision and if the single precision efficiency 
is computationally higher than that with double precision, the use of polynomial preconditioner 
has a chance to improve the linear solver timing.
With this motivation double precision (DP) solvers coupled with single-precision (SP) preconditioners 
(so called mixed precision solvers) have been proposed~\cite{mixedprec}.
Since the SP performance is much higher than DP for recent commodity architectures, such as 
IA32 or AMD64/Intel64 architecture, GPGPU, and Cell B.E, the mixed precision solver should be 
one of the choice to improve the algorithm.

The solver algorithm with the mixed precision (or flexible) preconditioner 
is based on the Richardson iteration (or iterative refinement) technique~\cite{Durr:2008rw}.
The solver requires ``flexibility'' which means that the preconditioner can be 
changed from iteration to iteration in the solver algorithm.
This is needed since the precision conversion between SP and DP cannot be expressed in a fixed matrix 
form in DP.
The iterative refinement algorithm to solve $Dx=b$ with a flexible preconditioner $P\sim D^{-1}$
is given in Alg.~\ref{alg:Richard}.
The preconditioner $P$ is computed at the 4th line and the accuracy can vary  
from iteration to iteration.
To compute the preconditioner $P$, any iterative solver can be used; this is called an inner-solver.
The residual and solution are updated at the 5th line. 
The accumulated residual and the solution hold the relation $r=b-Dx$ in DP.
Any flexible preconditioned solver can be constructed via the iterative refinement algorithm.
The mixed precision solver for lattice QCD has been used in 
Refs~\cite{sap+gcr,Luscher:2005rx,Durr:2008rw,PACSCSDDHMC} and a significant speed up 
has been observed. 

\begin{algorithm}
\caption{Richerdson iteration}
\label{alg:Rich}
  \begin{algorithmic}[1]
  \STATE $x$ is an initial guess, $P$ is an approximate inverse, $P\sim D^{-1}$ 
(preconditioner),
  \STATE Calculate initial residual $r = b - D x$,
  \LOOP
  \STATE Calculate correction vector $\delta x = P r \sim D^{-1}r$ 
         by roughly solving $D \delta x = r$.
  \STATE $r = r - D \delta x$,\ \ \ $x = x +   \delta x$,
  \IF{$||r||$ is small enough}
    \STATE break.
  \ENDIF
  \ENDLOOP
  \end{algorithmic}
  \label{alg:Richard}
\end{algorithm}

\paragraph{Preconditioner with spectrum information}
The solver performance depends on the spectrum and the condition number of 
the coefficient matrix. 
If the spectrum information is known the linear equation can be preconditioned.

Let $C_p=\{c_0,c_1,\cdots, c_{p-1}\}$ be the $p$-dimensional eigen-subspace basis
corresponding to the small eigenvalues of $D$. $C_p$ satisfies the partial Schur 
decomposition as
\begin{equation}
  D C_{p} = C_{p} T_{p}, \ \ \ \  C_{p}^\dag C_{p} = I_{p},
\end{equation}
where $T_{p}$ is a strictly upper triangular $p$-dimensional matrix and 
contains eigenvalues as the diagonal elements. 
There are two methods for preconditioning with the eigen-subspace.
One is an additive construction and another is a multiplicative one.
The additive preconditioner which removes the subspace from $D$ is called 
deflation~\cite{deflWilcox} and can be constructed as follows.
Using the projection operators $P$ and $Q$ defined by
\begin{equation}
    P = (1 - C_{p} C_{p}^{\dag}),\ \ \   Q = ( 1 - C_{p}T_p^{-1}C_{p}^{\dag}D),\ \ \   PD = DQ,
\label{eq:additive}
\end{equation}
we can remove the near zero eigen-subspace form $D$.
The equation $PD \tilde{x} = Pb$ is easily solved due to a reduced condition number and
the full solution is obtained by $x= Q \tilde{x} + C_{p}T^{-1}C^{\dag}_{p}b$.
The multiplicative preconditioner~\cite{AMGQCD} to reduce the condition number with spectrum 
information is 
\begin{equation}
    P = C_p T^{-1} C_p^{\dag}.
\label{eq:multiplicative}
\end{equation}
This mimics $D^{-1}$ in the subspace and the preconditioned equation $PD x = Pb$ 
is solved easily again.

The bottleneck to use the spectrum information as a preconditioner is 
the construction cost of the subspace. The cost reduction is achieved by several independent ways;
(i) use the approximate eigen-subspace to reduce the cost, because 
    exact eigen-subspace is not always required for the preconditioner.
(ii) reuse/recycle the subspace among multiple linear equations with different right-hand vectors.
(iii) construct subspace within the  Krylov-subspace iterative linear solver,
     because the eigen-subspace can be constructed from the same Krylov-subspace and
     the task for the linear solver and the eigen solver can be overlapped.
These methods are combined with deflation technique and a significant improvement 
has been observed.
The details for deflation technique and recent developments are described 
in Ref.~\cite{deflWilcox}.

\paragraph{Deflation with low-mode and local coherence}
The dimension of subspace affects the performance of the deflation preconditioner.
Since the density of near zero eigen-modes is proportional to the system size 
$V$ in order to keep the condition number at a lower level we have to increase 
the dimension as the volume. The cost to find an eigenmode is also proportionals to $V$.
This means that the total cost is of $O(V^2)$. 
This bottleneck can be circumvented by incorporating the physical insight on 
the near zero eigen-modes of $D$. 

L{\"u}scher has constructed a low-mode deflation subspace via a geometric lattice blocking
assuming {\it local coherence} property of the low modes~\cite{deflluscher}.
The subspace is constructed as follows;
(i) supply the low-mode enhanced vectors as
($\psi_{l}(x) \sim (D_{c}^{-1})^{\nu}\eta_{l}$, with random vectors 
$\eta_l$, $l=1,\cdots,N_s$), 
where $N_{s} \sim 12$ or $O(10)$, $\nu\sim 3$, 
and $D_{c}$ the Dirac operator near critical mass with rough approximate inversion,
(ii) block $\psi_{l}(x)$ as
\begin{equation}
  \psi_l^{\Lambda}(x) =
  \left\{
  \begin{array}{cc}
   \psi_{l}(x) &  \mbox{if $x \in \Lambda$},\\
             0 &  \mbox{if $x \not \in \Lambda$}. \\
  \end{array}
  \right.
\end{equation}
and then $\psi_l^{\Lambda}(x)$ are orthogonalized to yield $c_{l}^{\Lambda}(x)$.
The subspace is constructed as 
$C_p=\{ c_l^{\Lambda}(x), \Lambda = \mbox{lattice domain block index}\}$.

The dimension $p$ of $C_p$ is $p=N_s\times (\# \mbox{of blocks})$ and 
is significantly enlarged by the blocking.
This blocked subspace has a significant benefit that the memory requirement is low
compared to the full vector basis when compute node parallelism is matched 
to the domain decomposition. This removes the $O(V^2)$ bottleneck practically.
The low-mode approximation for $D$ leads to
\begin{equation}
    D_{\mathrm{low}} (i,\Lambda;j,\Lambda') = (c^{\Lambda}_{i})^{\dag} D c^{\Lambda'}_{j},
\end{equation}
where the index $\Lambda$ and $\Lambda'$ represent the coarse lattice grid index, and 
$i$ and $j$ $(=1,\cdots,N_s)$ mimic the internal degree of freedom similar to color-spin.
The low-mode $D_{\mathrm{low}}$ still has the nearest-neighbor structure on the coarse grid 
and is easy to invert via iterative solvers.
This blocking is very similar to real space RG-blocking. 

L{\"u}scher constructed the deflation preconditioner from this blocked subspace as
\begin{equation}
  P = (1 - DC_p (D_{\mathrm{low}})^{-1}C_p^{\dag}), \ \ \ 
  Q = (1 - C_p (D_{\mathrm{low}})^{-1}C_p^{\dag}D), \ \ \ 
  PD = D Q,
\end{equation}
and the solution for $Dx=b$ is obtained with 
$x=Q\tilde{x}+DC_p(D_{\mathrm{low}})^{-1}C_p^{\dag}b$
by solving $PD \tilde{x}=Pb$ for $\tilde{x}$.
A significant improvements and removal of critical slowing down at small quark masses
are observed not only in solver performance but also in the DDHMC 
algorithm~\cite{deflluscher}.

\paragraph{Multigrid with low-mode}
The RG-blocking like point of view for linear solvers is called the multigrid (MG) 
solver/preconditioner.
The MG iteration is based on the multiplicative preconditioner 
as Eq.~(\ref{eq:multiplicative}). The extension to multilevel MG is straightforward.
In the context of the MG algorithm the operators $C_p^{\dag}$ and $C_p$ correspond to 
the restriction operator and interpolation (prolongation) operator respectively. 
The efficiency of MG depends strongly on the choice of the subspace $C_p$.

The earlier attempt of the MG approach for lattice field theory was based on 
the purely geometric multigriding combined with the RG blocking for link and
fermion fields and mass parameters~\cite{NaiveMG}. 
Later the multigrid method has been generalized and arranged to 
the projective multigrid~\cite{ProjMG}.
In Refs.~\cite{ProjMG} the projection operator was constructed 
from the eigen-vectors of the {\it block-restricted} lattice Dirac operator 
$D(x,y):x,y \in \Lambda$. With this choice the enhancement of solver speed has been 
achieved at modest quark masses, while the critical slowing down near 
critical mass still remained.

Recently the adaptive multigrid method has been investigated and a significant 
speed up and removal of critical slowing down have been observed 
for the U(1) Schwinger model~\cite{AdaptiveMG}.
The key in this success is the same choice for 
projection/interpolation operator as  L\"{u}scher's for the deflation subspace.
The extension to 4D QCD has been presented this year by Clark {\it et al.}~\cite{AMGQCD}, 
where the relation between the deflation and adaptive MG methods is clearly explained
(like Eqs.~(\ref{eq:additive}) and (\ref{eq:multiplicative})).
The removal of critical slowing down is also observed for the QCD case with 
the adaptive MG method.

\section{Summary}
The lattice QCD simulations with dynamical quarks still require 
huge amount of computational resources and development of efficient numerical algorithms.
The recent trend in computational machines in these years is the many core architecture. 
The peak performance is rapidly growing while the growth ratio of memory bandwidth 
is rather slow; the gap between the two is still the primary obstacle for high performance computing.
To fill the gap the memory subsystem hierarchy is becoming increasingly complicated through  an 
insertion an intermediate cache system or local memory attached to cores.
The algorithm and programming close to the hardware implementation is important
to keep high sustained performance for lattice QCD simulations.
GPGPU computing for lattice QCD is becoming more familiar by the appearance 
of the GPGPU-dedicated general programming language. 
Attempts are starting to make a large scale simulation with high cost performance, 
but the bottleneck is again the bandwidth between the GPU card and the host computer.
The mixed precision/flexible preconditioner technique is a possible solution to
remove the difficulty by reducing data flow.

Preconditioning of the QCD partition function and the Sexton-Weingarten multiple-time scale MD 
is now a common and accepted technique for dynamical lattice QCD simulations. 
In this talk I covered several preconditioning techniques. 
The polynomial preconditioning and the RHMC algorithm represent a general methodology  
independent of the details of quark actions, 
while the geometric one strongly depends on the structure.
With the L\"{u}scher's DDHMC algorithm, which fully makes use of the nearest-neighbor 
coupling nature of the Wilson quark action, simulations near the chiral limit is now possible.
The geometric preconditioner dedicated and customized for a given quark action 
will further enhances the performance of the HMC algorithm for other quark actions.
For the MD integrator issue, the tuning method via the parametrized Omelyan integrator 
has been developed and is now available.

The solver performance has been improved via the deflation technique
which removes the critical slowing down and solver stagnation problems.
The local coherence property of the low modes further enhances the performance.
This year an adaptive multigrid technique has been applied for QCD and shown to remove 
the critical slowing down. 
The adaptive multigrid also makes use of the local coherence property of low modes in 
a different way.
The realization of the local coherence deflation and the adaptive multigrid 
strongly depends on the structure of the quark action
 (nearest-neighbor coupling action) again.

Combining these improvements the dynamical lattice QCD simulations are now possible
close to the chiral limit on lattices with the lattice spacing $a\sim 0.1$fm and the 
volume $L=3$fm for Wilson type and staggered type fermions using $O(10)$ TFlops supercomputers.
The empirical cost performance formula for the DDHMC algorithm~\cite{DDHMClight} 
indicates the possibility of the dynamical QCD simulations at:
  (i) $m_q\sim 5$ MeV, $L\sim 6$ fm, $a\sim 0.1$ fm,
 (ii) $m_q\sim 5$ MeV, $L\sim 2$ fm, $a\sim 0.03$ fm,
with future 10 Peta flops supercomputers. 
The first case can handle multiple hadrons in a single box and the second one 
the charm quark physics with dynamical charm. 
These problems contain one more physical scales, 
(i) pion wave length $1/m_{\pi}$ and (ii) charm quark wave length $1/m_{c}$
in addition to $1/L$  and $1/a$.
Numerical simulations for multiple physical scale problems are difficult in general. 
A multiple data blocking is required when the problem is treated with 
a fat-node supercomputer.
A matching between the physical scale and the machine structure is recommended 
for this case.
I expect that the algorithm with fully utilizing the data blocking structure and
physical scale blocking structure will cover these problems,
for exapmle nesting the domain-decomposed HMC, or nesting deflation and multigrid via local coherence.
Attempts at these physics have already started by several collaborations
using $O(100)$ TFlops machines by making use of the improvement approaches described 
in the paper.

\begin{acknowledgments}
I would like to thank all the colleagues who made their results available 
before the conference.
I also thank to A.~Kennedy, M.~Clark, W.~Wilcox for valuable discussions and 
comments on the improvement methods and algorithms, and M.~Okawa and A.~Ukawa
for proof reading on the manuscript.
This work is supported in part by the Grants-in-Aid
of Ministry of Education (Nos. 20740139, 18104005).
\end{acknowledgments}



\begin{thebibliography}{99}
\bibitem{Duane:1987de}
  S.~Duane, A.~D.~Kennedy, B.~J.~Pendleton and D.~Roweth,
    Phys.\ Lett.\  B {\bf 195} (1987) 216.
  

\bibitem{Sexton:1992nu}
  J.~C.~Sexton and D.~H.~Weingarten,
    Nucl.\ Phys.\  B {\bf 380} (1992) 665.
  

\bibitem{deForcrand:1996ck}
  P.~de Forcrand and T.~Takaishi,
    Nucl.\ Phys.\ Proc.\ Suppl.\  {\bf 53} (1997) 968
  [arXiv:hep-lat/9608093].
  

\bibitem{deflWilcox}
For recent review on the deflation technique for lattice QCD, see:
W.~M.~Wilcox,
    \pos{PoS(LATTICE 2007)025};
   and references there in.

\bibitem{deflluscher}
  M.~L{\"u}scher,
  JHEP {\bf 0712}, 011 (2007);
  JHEP {\bf 0707}, 081 (2007).

\bibitem{AdaptiveMG}
  J.~Brannick, R.~C.~Brower, M.~A.~Clark, J.~C.~Osborn and C.~Rebbi,
  Phys.\ Rev.\ Lett.\  {\bf 100} (2008) 041601
  [arXiv:0707.4018 [hep-lat]];
  \pos{PoS(LATTICE 2007)029}
  [arXiv:0710.3612 [hep-lat]].

\bibitem{Holmgren:2004nk}
  D.~J.~Holmgren,
    Nucl.\ Phys.\ Proc.\ Suppl.\  {\bf 140}, 183 (2005)
  [arXiv:hep-lat/0410049].
  

\bibitem{Wettig:2005zc}
  T.~Wettig,
  \pos{PoS(LAT2005)019}
  [arXiv:hep-lat/0509103].
  

\bibitem{ClearSpeed}
  ClearSpeed Web page, [http://www.clearspeed.com/].

\bibitem{Ibrahim:2008zz}
  K.~Ibrahim {\it et al.},
    arXiv:0808.0391 [hep-lat].

\bibitem{Belletti:2007pp}
  F.~Belletti {\it et al.},
    \pos{PoS(LATTICE 2007)039}
  [arXiv:0710.2442 [hep-lat]].
  

\bibitem{Nakamura:2006zzc}
  A.~Nakamura and S.~Motoki,
    \pos{PoS(LATTICE 2007)040}.
  

\bibitem{Spray:2008nt}
  J.~Spray, J.~Hill and A.~Trew,
      arXiv:0804.3654 [hep-lat].
  

\bibitem{QPACE}
  H.~Baier {\it et al.},
    arXiv:0810.1559 [hep-lat],  in these proceedings.

\bibitem{CUDAZONE} Nvidia CUDA ZONE, \verb+[http://www.nvidia.co.jp/object/cuda_home_jp.html]+.

\bibitem{Firestream} AMD Stream Computing, 
\verb+[http://ati.amd.com/technology/streamcomputing/index.html]+.

\bibitem{GPGPUQCD}
  G.~I.~Egri, Z.~Fodor, C.~Hoelbling, S.~D.~Katz, D.~Nogradi and K.~K.~Szabo,
  ``Lattice QCD as a video game,''
  Comput.\ Phys.\ Commun.\  {\bf 177} (2007) 631
  [arXiv:hep-lat/0611022].

\bibitem{Fodor:2007ue}
  Z.~Fodor,
    arXiv:0712.2930 [hep-lat].
  

\bibitem{RenzoCUDA}
Francesco~Di~Renzo, ``GPU computing for 2-d spin systems: CUDA vs OpenGL'', in these proceeings.

\bibitem{RebbiCUDA}
  K.~Barros, R.~Babich, R.~Brower, M.~A.~Clark and C.~Rebbi,
   PoS {\bf LATTICE2008} (2008) 045
  [arXiv:0810.5365 [hep-lat]], in these proceeings.
  

\bibitem{TESLA} Nvidia TESLA series, \verb+[http://www.nvidia.com/object/tesla_computing_solutions.html]+

\bibitem{PetAPE}
Davide Rossett (INFN APE Group), private communication, to appear in Nuovo Cimento.

\bibitem{Roadrunner}
  Roadrunner Home
  [http://www.lanl.gov/roadrunner/index.shtml].

\bibitem{TopFH}
  TOP 500 [http://www.top500.org/system/9485].

\bibitem{Oyanagi}
Y.~Oyanagi, Comput. Phys. Commun. {\bf 42}, 333 (1986).

\bibitem{SSOR}
  S.~Fischer, A.~Frommer, U.~Glassner, T.~Lippert, G.~Ritzenhofer and K.~Schilling,
    Comput.\ Phys.\ Commun.\  {\bf 98} (1996) 20
  [arXiv:hep-lat/9602019].
  

\bibitem{Hasenbusch}
  M.~Hasenbusch,
  Phys.\ Lett.\  B {\bf 519} (2001) 177
  [arXiv:hep-lat/0107019];
  \pos{PoS(LAT2005)116}
  [arXiv:hep-lat/0509080];
  M.~Hasenbusch and K.~Jansen,
    Nucl.\ Phys.\  B {\bf 659} (2003) 299
  [arXiv:hep-lat/0211042];
  Nucl.\ Phys.\ Proc.\ Suppl.\  {\bf 119} (2003) 982
  [arXiv:hep-lat/0210036];
  Nucl.\ Phys.\ Proc.\ Suppl.\  {\bf 106} (2002) 1076
  [arXiv:hep-lat/0110180];

\bibitem{Urbach:2005ji}
  C.~Urbach, K.~Jansen, A.~Shindler and U.~Wenger,
      Comput.\ Phys.\ Commun.\  {\bf 174} (2006) 87
  [arXiv:hep-lat/0506011].
  

\bibitem{AliKhan:2003br}
  A.~Ali Khan {\it et al.}  [QCDSF Collaboration],
    Phys.\ Lett.\  B {\bf 564} (2003) 235
  [arXiv:hep-lat/0303026].
  

\bibitem{Gockeler:2007rm}  M.~Gockeler {\it et al.}  [QCDSF Collaboration],
   \pos{PoS(LATTICE 2007)041}
  [arXiv:0712.3525 [hep-lat]].
  

\bibitem{Peardon:2002wb}
  M.~J.~Peardon and J.~Sexton  [TrinLat Collaboration],
      Nucl.\ Phys.\ Proc.\ Suppl.\  {\bf 119} (2003) 985
  [arXiv:hep-lat/0209037].
  

\bibitem{Alexandrou:1999ii}  
    C.~Alexandrou, P.~de Forcrand, M.~D'Elia and H.~Panagopoulos,
    Phys.\ Rev.\ D {\bf 61} (2000) 074503
  [arXiv:hep-lat/9906029]; 
    Nucl.\ Phys.\ Proc.\ Suppl.\  {\bf 83} (2000) 765
  [arXiv:hep-lat/9909004];
    P.~de Forcrand,
    Nucl.\ Phys.\ Proc.\ Suppl.\  {\bf 73} (1999) 822
  [arXiv:hep-lat/9809145].

\bibitem{Luscher:1993xx}   M.~L\"{u}scher,
      Nucl.\ Phys.\ B {\bf 418} (1994) 637
  [arXiv:hep-lat/9311007].

\bibitem{Ishikawa:2006pb}
  K.~I.~Ishikawa {\it et al.}  [PACS-CS Collaboration],
   \pos{PoS(LAT2006)027}
  [arXiv:hep-lat/0610037].
  

\bibitem{UVsuppressF}
  A.~Borici,
    Nucl.\ Phys.\ Proc.\ Suppl.\  {\bf 119} (2003) 988
  [arXiv:hep-lat/0208048];
  A.~Borici  [UKQCD Collaboration],
    J.\ Comput.\ Phys.\  {\bf 189} (2003) 454
  [arXiv:hep-lat/0208034];
  A.~Borici  [UKQCD collaboration],
    Phys.\ Rev.\  D {\bf 67} (2003) 114501
  [arXiv:hep-lat/0205011].
  

\bibitem{Peardon:2000si}
  M.~J.~Peardon,
    arXiv:hep-lat/0011080.
  

\bibitem{Ordering}
S. Doi and A. Lichnewsky, 
``A graph-theory approach for analyzing the effects of ordering on ILU preconditioning'', 
INRIA report 1452 (1991).

\bibitem{Luscher:2005rx}
  M.~Luscher,
    Comput.\ Phys.\ Commun.\  {\bf 165} (2005) 199
  [arXiv:hep-lat/0409106].
  

\bibitem{DDHMClight}
  L.~Del Debbio, L.~Giusti, M.~Luscher, R.~Petronzio and N.~Tantalo,
      JHEP {\bf 0702} (2007) 056
  [arXiv:hep-lat/0610059];
  L.~Del Debbio, L.~Giusti, M.~Luscher, R.~Petronzio and N.~Tantalo,
      JHEP {\bf 0702} (2007) 082
  [arXiv:hep-lat/0701009].
  

\bibitem{PACSCSDDHMC}
  S.~Aoki {\it et al.}  [PACS-CS Collaboration],
    arXiv:0807.1661 [hep-lat];
  P.~ .~Ukita {\it et al.}  [PACS-CS Collaboration],
    arXiv:0810.0563 [hep-lat];
  P.~:.~Kadoh {\it et al.}  [PACS-CS Collaboration],
      arXiv:0810.0351 [hep-lat];
  Y.~Kuramashi {\it et al.}  [PACS-CS Collaboration],
   \pos{PoS(LAT2006)029}
  [arXiv:hep-lat/0610063].
  

\bibitem{Hasenbusch:2007er}
  M.~Hasenbusch,
      \pos{PoS(LATTICE 2007)033}
  [arXiv:0710.0066 [hep-lat]].
  

\bibitem{Borici:2007bp}
  A.~Borici,
    \pos{PoS(LATTICE 2007)065}
  [arXiv:0711.0508 [hep-lat]];
      arXiv:0704.2341 [hep-lat].
  

\bibitem{RHMC}
  M.~A.~Clark and A.~D.~Kennedy,
      Phys.\ Rev.\  D {\bf 75} (2007) 011502
  [arXiv:hep-lat/0610047];
      Nucl.\ Phys.\ Proc.\ Suppl.\  {\bf 140} (2005) 838
  [arXiv:hep-lat/0409134];
  M.~A.~Clark and A.~D.~Kennedy,
    Nucl.\ Phys.\ Proc.\ Suppl.\  {\bf 129} (2004) 850
  [arXiv:hep-lat/0309084];
  A.~D.~Kennedy, I.~Horvath and S.~Sint,
      Nucl.\ Phys.\ Proc.\ Suppl.\  {\bf 73} (1999) 834
  [arXiv:hep-lat/9809092].
  

\bibitem{RHMC2}
  M.~A.~Clark and A.~D.~Kennedy,
      Phys.\ Rev.\ Lett.\  {\bf 98} (2007) 051601
  [arXiv:hep-lat/0608015];
    M.~A.~Clark, Ph.~de Forcrand and A.~D.~Kennedy,
    \pos{PoS(LAT2005)115}
  [arXiv:hep-lat/0510004];
    M.~A.~Clark,
    \pos{PoS(LAT2006)004}
  [arXiv:hep-lat/0610048].
  

\bibitem{Kennedy:2004tj}
  A.~D.~Kennedy,
    Nucl.\ Phys.\ Proc.\ Suppl.\  {\bf 128C} (2004) 107
  [arXiv:hep-lat/0402037].
  

\bibitem{Kennedy:2006ax}
  A.~D.~Kennedy,
    arXiv:hep-lat/0607038.
  

\bibitem{Kogut:2006jg}
  J.~B.~Kogut and D.~K.~Sinclair,
    Phys.\ Rev.\  D {\bf 74} (2006) 114505
  [arXiv:hep-lat/0608017].
  

\bibitem{Durr:2008rw}  S.~Durr {\it et al.},
    arXiv:0802.2706 [hep-lat].
  

\bibitem{Aoki:2005vt}  Y.~Aoki, Z.~Fodor, S.~D.~Katz and K.~K.~Szabo,
      JHEP {\bf 0601} (2006) 089
  [arXiv:hep-lat/0510084].
  

\bibitem{Antonio:2008zz}  D.~J.~Antonio {\it et al.}  [RBC Collaboration and UKQCD Collaboration],
    Phys.\ Rev.\  D {\bf 77} (2008) 014509
  [arXiv:0705.2340 [hep-lat]];
  C.~Allton {\it et al.}  [RBC and UKQCD Collaborations],
      Phys.\ Rev.\  D {\bf 76} (2007) 014504
  [arXiv:hep-lat/0701013];
  D.~J.~Antonio {\it et al.}  [RBC and UKQCD Collaborations],
      Phys.\ Rev.\  D {\bf 75} (2007) 114501
  [arXiv:hep-lat/0612005].
  

\bibitem{Takaishi:2007ga}  T.~Takaishi and A.~Nakamura,
    \pos{PoS(LATTICE 2007)229}
  [arXiv:0711.3888 [hep-lat]].
  

\bibitem{Borici:1995am}
  A.~Borici and P.~de Forcrand,
    Nucl.\ Phys.\  B {\bf 454} (1995) 645
  [arXiv:hep-lat/9505021].
  

\bibitem{Takaishi:2001um}
  T.~Takaishi and P.~de Forcrand,
    Int.\ J.\ Mod.\ Phys.\  C {\bf 13} (2002) 343
  [arXiv:hep-lat/0108012].
  

\bibitem{Aoki:2001pt}
  S.~Aoki {\it et al.}  [JLQCD Collaboration],
      Phys.\ Rev.\  D {\bf 65} (2002) 094507
  [arXiv:hep-lat/0112051].

\bibitem{Oliver}
  O.~Witzel,
    arXiv:0809.1010 [hep-lat] in these proceedings.
  

\bibitem{Contourinteg}
Nicholas Hale and Lloyd N. Trefethen), SIAM J. Numer. Anal. 46 (5):2505-2523, 2008.

\bibitem{Omelyan}
  I.~P.~Omelyan, I.~M.~Mryglod, and R.~Folk, Comput.~Phys.~Commun. {\b 151} (2003) 272;
  Phys.~Rev.\ E {\bf 65} (2002) 056706.

\bibitem{Takaishi:2005tz}
  T.~Takaishi and P.~de Forcrand,
      Phys.\ Rev.\  E {\bf 73} (2006) 036706
  [arXiv:hep-lat/0505020].
  

\bibitem{Kennedy:2007cb}
  A.~D.~Kennedy and M.~A.~Clark,
    \pos{PoS(LATTICE 2007)038}
  [arXiv:0710.3611 [hep-lat]].
  

\bibitem{HMCPoisson}
  M.~A.~Clark, A.~D.~Kennedy and P.~J.~Silva,
  arXiv:0810.1315 [hep-lat], in theese proceedings.

\bibitem{CreutzGockesch} 
  M.~Creutz and A.~Gockesh, Phys.~Rev.~Lett. {\bf 63} (1989) 9.

\bibitem{CampostriniRossi}
  M.~Campostrini and P.~Rossi, Nucl.~Phys.~{\bf B} 329 (1990) 753.

\bibitem{Yoshida}
  H.~Yoshida, Phys.~Lett.~A {\bf 150} (1990) 262.

\bibitem{Suzuki}
  M.~Suzuki, Phys.~lett.~A {\bf 146} (1990) 319.

\bibitem{Takaishi:1999bi}
  T.~Takaishi,
    Comput.\ Phys.\ Commun.\  {\bf 133} (2000) 6
  [arXiv:hep-lat/9909134].
  
  

\bibitem{Boyle:2007fn}
  P.~Boyle  [RBC Collaboration and UKQCD Collaboration],
      \pos{PoS(LATTICE 2007)005}
  [arXiv:0710.5880 [hep-lat]].
  

\bibitem{Boucaud:2008xu}
  Ph.~Boucaud {\it et al.}  [ETM collaboration],
      arXiv:0803.0224 [hep-lat].
  

\bibitem{mixedprec}
A.~Buttari, J.~Dongarra, J.~Kurzak, P.~Luszczek, and S.~ Tomov,
``Using mixed precision for sparse matrix computations to 
enhance the performance while achieving 64-bit accuracy'',
to appear in ACM Trans.~Math.~Soft.

\bibitem{sap+gcr}
M. L\"uscher, Comput. Phys. Commun. {\bf 156}, 209 (2004).

\bibitem{AMGQCD}
 M.~Clark {\it et al.}, ``The removal of critical slowing down'', in these proceedings. 

\bibitem{NaiveMG}
  R.~C.~Brower, E.~Myers, C.~Rebbi and K.~J.~M.~Moriarty,
  in {\it Multigrid Methods: Theory, Applications and Supercomputing},
  Proceedings of the Third Copper Mountain Conference on Multigrid Methods, 
  Copper Mountain, Colorado, 1987, edited by S.~F.~McCormick (Marcel Dekker, 1988, New York) 85;    
  P.~G.~Lauwers and T.~Wittlich,
  Int.\ J.\ Mod.\ Phys.\  C {\bf 4} (1993) 609;
  Nucl.\ Phys.\ Proc.\ Suppl.\  {\bf 30} (1993) 261.

\bibitem{ProjMG}
  R.~C.~Brower, R.~G.~Edwards, C.~Rebbi and E.~Vicari,
  Nucl.\ Phys.\  B {\bf 366} (1991) 689;
  R.~C.~Brower, C.~Rebbi and E.~Vicari, 
  Phys.~Rev.~D {\bf 48} (1991) 1965;
  T.~Kalkreuter, Phys.~Rev.~D {\bf 48} (1993) R1926;
  Phys.~Rev.~D {\bf 51} (1995) 1305.


\end{thebibliography}
\end{document}